\documentclass{appolb}
\usepackage{graphicx}
\usepackage{amsmath}

\begin{document}
\title{DISCRETE SYMMETRIES CP, T, CPT
\thanks{"Jagiellonian Symposium on Fundamental and Applied Subatomic Physics" in Krakow}%
}
\author{Jos\'e Bernab\'eu \\\address {Departament of Theoretical Physics, University of Valencia \\\& IFIC
Joint Centre Univ.Valencia--CSIC,\\ C/Catedr\'atico Jos\'e Beltr\'an 2, E-46980 Paterna, Spain, \\jose.bernabeu@uv.es}}

\maketitle
\begin{abstract}
The role of Symmetry Breaking mechanisms to search for New Physics is of highest importance. We discuss the status and prospects of the Discrete Symmetries CP, T, CPT looking for their separate Violation in LHC experiments and meson factories.
\end{abstract}

\section{Introduction}
Symmetries have been an essential ingredient in the understanding of the physical laws of Nature.The assumption of the form invariance of the dynamical equations under a symmetry transformation of the physical magnitudes leads to observable consequences, with regularities, conservation laws and invariant observables that act as guiding components for the dynamics. Even more interesting, Symmetry Breaking through a definite mechanism is a source of new phenomena and new physics. In this talk I will concentrate on the Discrete Symmetries CP, T and CPT.
\\In Section 2 we discuss that the current level of experimental accuracy and theoretical uncertainties leaves room for additional sources of CP-Violation (CPV) beyond the Cabibbo-Kobayashi-Maskawa (CKM) mixing mechanism, as required by Baryogenesis in the Universe. We identify potential transitions in B-physics, using Flavour-Changing-Neutral-Current (FCNC) processes and CPV asymmetries, able to incorporate virtual contributions of New Physics.  
\\Section 3 is devoted to TRV concepts and results: what is “Time Reversal” in classical and quantum mechanics, the NO-GO argument for its search with unstable particles, its By-pass using Entanglement and the Decays as Filtering Measurements for entangled meson systems in the B-Factories and in the $\Phi$-Factory, the role of time-ordered decay channels to Flavour and CP-eigenstates for disentangling genuine separate independent asymmetries for CP, T and CPT and the 14$\sigma$ observation of TRV by the BABAR experiment. 
Section 4 emphasizes the interest in the search of CPTV for transitions and not only in the expectation values of masses and lifetimes of elementary particles. We distinguish explicit CPT symmetry breaking mechanisms from physical scenarios in which the CPT-operator implementing the symmetry is ill-defined. For entangled systems this second alternative leads to the $\omega$-effect, a component of the wrong symmetry in the coherent state of neutral mesons. Finally, Section 5 presents our results, conclusions and prospects.
\section{CP-Violation}
The now well established CPV in the quark sector can be successfully accommodated within the Standard Model (SM) of particles and fields through the CKM quark-mixing [1]. For three families, the unitarity conditions lead to triangles like the bd-triangle of Fig.1.

\begin{figure}
\centerline{%
\includegraphics[width=8.5cm]{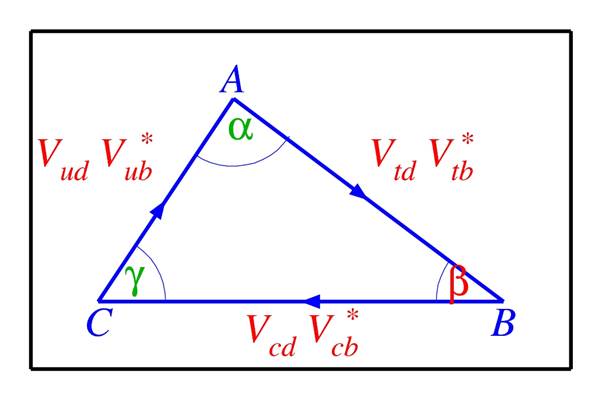}}
\caption{The bd unitarity triangle representing the CKM mixing}
\end{figure}

Since the three sides are of comparable length the angles are sizeable and one expects large CP asymmetries in Bd decays in the SM. There are other two triangles which almost collapse to a line. This gives an intuitive understanding of why CPV is small in the leading K decays (ds triangle) and in the leading Bs decays (bs triangle). Extensive tests of the CKM mechanism using experimental data show a high degree of consistency [2]. Two B factory colliders, PEP-II at SLAC in California and KEKB at KEK in Japan, with their corresponding detectors, BABAR [3] and Belle [4], have been operating in the last decade. We now have the LHCb experiment [5] at CERN with new and complementary information about rare decays,FCNC processes and CPV. For the Bd system CPV in the interference of mixing and no-mixing amplitudes for $B^{0}$ decays is observed in decay products which are accessible by both $B^{0}$ and $\bar{B^{0}}$. The CPV Asymmetry can be written as 

\begin{equation}
A_f (t)= S_f sin(\Delta mt) - C_f  cos(\Delta mt)
\end{equation}
where $\Delta \Gamma$= 0 is assumed. The most precise asymmetries are measured in the tree-dominated b$\rightarrow$ ccs transitions, such as $B^{0}\rightarrow\psi$ KS, and are given by [6] S = + 0.682 $\pm$ 0.019. The penguin contributions are very small, so that one has the interpretation in the SM  S = sin(2$\beta$), C=0. CPV in the interference of mixing and decay in the $B^{0}\rightarrow\pi^{+}\pi^{-}$ mode is given by [6]   S = $- 0.66 \pm$ 0.06. It is interpreted in the SM as S = sin(2$\alpha$).For the $B^{0}$system, the phase of the mixing amplitude is determined from the intermediate top-quark exchange in the box diagram, so that the SM interpretations of these interferences in terms of the $\beta$ and $\alpha$ CP angles of the unitarity triangle are apparent. On the contrary, the phase $\gamma$ in the unitarity triangle involves the interference of the sides for decays with charm and up quark constituents, without any relation to the mixing. The CP angle $\gamma$ is thus a measure of Direct CPV. Its measurement has been undertaken by BABAR, BELLE and LHCb with the decay $B^{+}\rightarrow DK^{+}$ and other related transitions. The extraction of $\gamma$ needs a detailed analysis involving in addition the presence of strong phases associated to final-state hadronic interactions. The present average value [7] is $\gamma = (67 + - 12)^{o}$. An ideal experiment would be one in B factories using Entanglement and detecting the pair of decays $\rightarrow\psi K^{0}$ and $\rightarrow\pi \pi$ at equal times without any effect of the mixing phase. It remains to be seen whether this gedanken experiment can become a real experiment in the upgraded SuperBELLE.
\\The rare decay $Bs\rightarrow\mu\mu$ has been observed by LHCb and CMS. Based on the presence of a FCNC penguin amplitude induced by Z-exchange as seen in the diagram 

\includegraphics[scale=0.50]{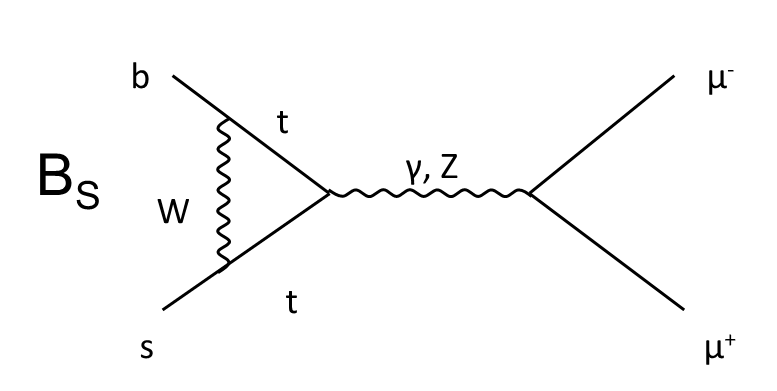}                         

one could expect a priori New Physics virtual effects induced by Non-Decoupling of longitudinal contributions. The result [8] presented in figure 2 has represented the latest disappointment of the scientific community. The experimental Branching Ratio is 

\begin{figure}
\centerline{%
\includegraphics[width=12cm]{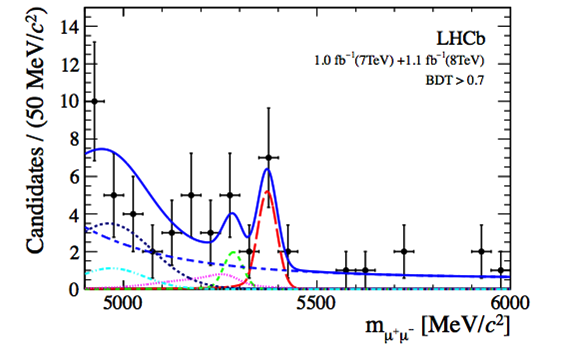}}
\caption{The invariant mass distribution of the $\mu\mu$ system showing the Bs-peak}
\end{figure}

\begin{equation}
B (B _s^0\rightarrow\mu^+ \mu^- )=(〖3.2〗 _{-1.2}^{+1.4} (stat) _{-0.3}^{+0.5} (syst))x 〖10〗^{-9}
\end{equation}                     

in perfect agreement with the SM value $(3.23 \pm 0.27)x10^{-9}$. 
In some models, like SuperSymmetry, the result could have been orders of magnitude different from the SM value. One has to be aware, however, that in this process the FCNC penguin is projected to a pseudoscalar, so that its contraction with the leptonic vertex leads to helicity suppression and an amplitude proportional to the mass of the lepton. The FCNC b-s penguin vertex can be probed under more general conditions by searching for the processes $B\rightarrow  K^{*} l^{+} l^{-}$, which opens new effective current operators in scalar-vector matrix elements. In fact, the process has been observed and the analysis, using the experimental results of LHCb making use of the OPE formalism, leads to intriguing tensions with the SM expectations in some of the effective operators [9]. If this discrepancy is associated to new physics longitudinal amplitudes of the mediators, one should seriously consider the search of the process  with $\nu\nu$, mediated by the Z, replacing $\mu\mu$.

\section{Time Reversal}
The symmetry transformation that changes a physical system with a given sense of the time evolution into another with the opposite sense is called Time-Reversal T. The T-transformation is implemented in the space of quantum states by the antiunitary operator $\textit{U}_T$ For a Hamiltonian H invariant under time reversal, [H;$\textit{U}_T$] = 0, the time-evolution operator \textbf{U}(t;t0) transforms as

\begin{equation}
U_T \textbf{U}(t,t_0)U_T^\dagger† =\textbf{U}^\dagger(t,t_0)
\end{equation}

The antiunitary character of $\textit{U}_T$ allows to write $\textit{U}_T$ = UK, where U is unitary ($U^{-1}=U^{\dagger}$) and K is an operator which complex conjugates all complex numbers. For the matrix elements of time-dependent transitions we have 

\begin{eqnarray}
\langle f \vert \textbf{U}(t,t_0)\vert i \rangle = \langle f \vert U_ T^\dagger U_T \textbf{U}(t,t_0) U_T^\dagger U_T \vert i \rangle =\nonumber
\\〈U_T f \vert \textbf{U}^\dagger (t,t_0 )\vert U_T i\rangle^*=  \langle U_T i\vert \textbf{U}(t,t_0)\vert U_T f \rangle
\end{eqnarray}

where time-reversal invariance is assumed. As a consequence, the comparison between $i\rightarrow f$ and $\textit{U}_T f \rightarrow \textit{U}_Ti$ transitions is a genuine test of this invariance.  
\\A direct evidence for TRV would mean an experiment which, by itself, is able to establish a non-vanishing genuine TRV asymmetry independent of CPV or CPT invariance. The problem is then the filtering of definite initial and final states of the neutral meson for the Reference and T-reverse transitions, something impossible for decaying particles. The solution [10-14] arises from the quantum mechanical properties imposed by the EPR entanglement [15, 16] between the two neutral B mesons produced in the Y(4S) resonance decay. Let us suppose that the Reference transition is defined by the time-ordered decay channels $l^+$ first, J/$\Psi$ $K_S$ later, as shown in the left-hand side of figure 3. The use of Entanglement plus the Decay as a Filtering Measurement tells us that the meson transition corresponds to $\bar{B^{0}}\rightarrow B_{-}$ . In terms of meson states, the T-reverse transition is then $B_-\rightarrow\bar{B_{0}}$ and the question arises: Which are the time-ordered decay channels which correspond to this T-reverse transition? For definite flavour and CP eigenstates, orthogonality of ${B^{0}},\bar{B^{0}}$ and $ B_+, B_-$ provides the solution: J/$\Psi$ KL first, $l^-$ later.

\begin{figure}
\centerline{
\includegraphics[scale=0.5]{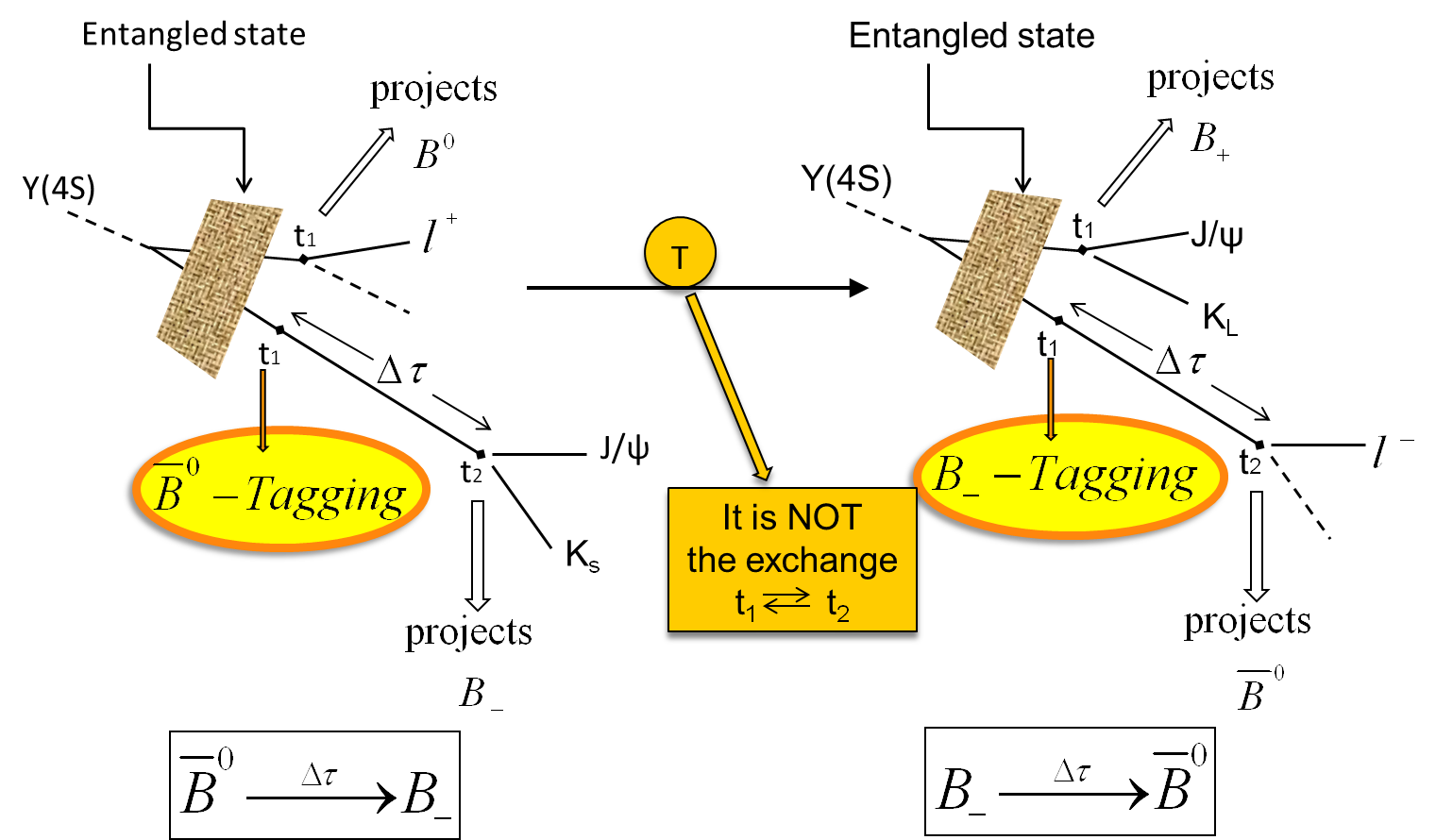}} 
\caption{Basic concepts explaining the preparation and detection of initial and final meson states, for the Reference and T-reverse transitions, by means of the esperimental time-ordered decay channels to definite flavour and CP eigenstates.}
\end{figure}

There are 8 experimentally independent Intensities for this kind of pairs of decay channels: 2 for Flavour X 2 for CP X 2 for the time-ordering of the decay channels. The four TRV independent raw symmetries measured by the BABAR experiment [17] are given in figure 4.

\begin{figure}
\centerline{
\includegraphics[scale=0.4]{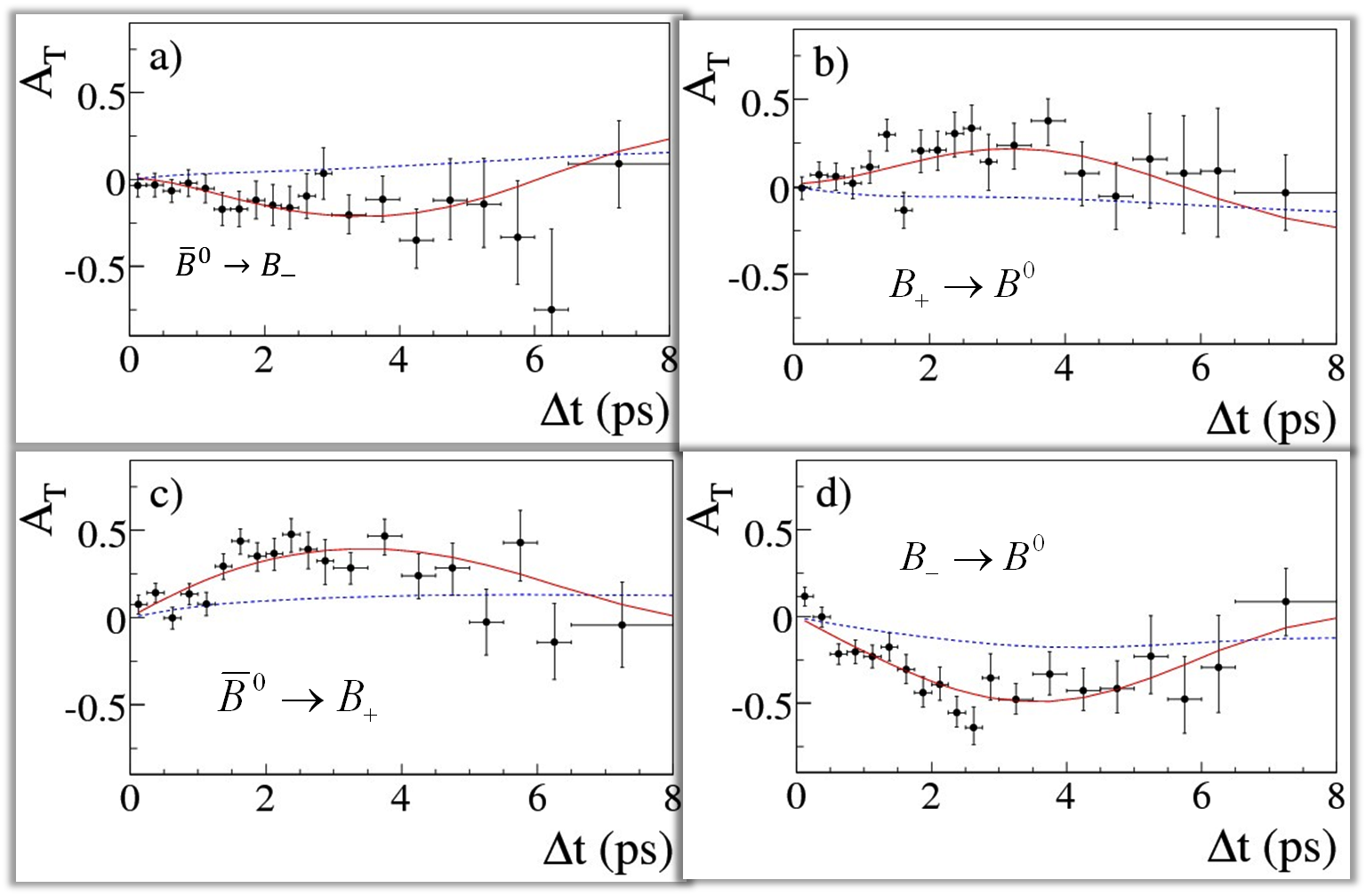}}
\caption{Experimental TRV asymmetries as function of $\Delta$t}   
\end{figure}

They correspond to the asymmetries between the Reference transitions $\bar{B^{0}}\rightarrow B_{-},B_{+}\rightarrow B^{0}$, $\bar{B^{0}}\rightarrow B_{+}, B_{-}\rightarrow B^{0}$, and their T-reverse transitions. The $\Delta$t dependence demonstrates that the TRV effect is built in the time evolution of the neutral B. The combined significance of the four measured nonvanishing asymmetries provides a conclusive 14$\sigma$ TRV effect.

\section{CPT Invariance}
The CPT theorem [19] establishes that interactions described by a Local  Quantum Field Theory, with Lorentz invariance and Hermiticity, are CPT invariant. The simplest tests of CPT invariance are the equality of masses and lifetimes for a particle and its antiparticle. 
The most precise test comes from the mass difference between $K^{0}$ and $\bar{K^{0}}$ [20].
It is very important to move the tests of CPT invariance beyond the comparison of diagonal terms for particle and antiparticle by considering these searches for transitions. For entangled systems, the proposed tests of separate CP, T, CPT symmetries in the neutral meson systems are based on the EPR-Entanglement existing in the Meson Factories as a consequence of Particle Identity:      $K^{0}$ and $ \bar{K^{0}}$ are two states of identical particles, connected by CPT. Besides the permutation operation $\textit{P}$ for space-time properties, the strangeness charge connection is made by C, so that for bosons the indistinguishibility requirement is C $\textit{P}$ = +.
\\In neutral meson factories, $K^{0},\bar{K^{0}}$ are produced by $\Phi-decay$ with J=1, S=0. This implies L=1 and C= -, so that $\textit{P}$ = -, i.e., an antisymmetric wave function. This antisymmetry is responsible for preserving $K^{0} \, \bar{K^{0}}$ terms only in the time evolution of the two-body system, including the Mixing . Similarly for orthogonal $B_{+} B_{-}$ terms only. This correlation is perfect for tagging: Flavour-Tag, CPTag [21].
\\The question is [22]: What if the Particle Identity is lost? In this case, the two particle system would not satisfy the requirement C $\textit{P}$ = +. In perturbation theory, if still J=1 with C=-, this breaking leads to a mixing of the “forbidden”  $\textit{P}$ = + symmetric state into the "allowed" $\textit{P}$ = - antisymmetricstate. This perturbative mixing is the 
$\omega$-effect: In the time  evolution of the system one finds now $\omega  K^{0} K^{0}$, $\omega$ $K_{+} K_{+}$ terms,..., i.e., a Demise of Tagging.
\\The decoherence implied by the $\omega$-effect is best seen for equal decay channels at $\Delta$ t = 0. For the ($\pi^{+}\pi^{-}$,$\pi^{+} \pi^{-}$) channel, the most prominent effect is the breaking of the Intensity I($\Delta$ t)$\sim$ 0 for small values of $\Delta$t, a result that was a consequence of the particle identity anti-correlation: no identical states at t1=t2. The KLOE experiment has obtained the first measurement of the $\omega$-parameter [23]:

\begin{equation} 
\left. 
\begin{matrix}
\text{Re}(\omega)=
\left(-1.6 
\begin{smallmatrix}+3.0\\ -2.1\end{smallmatrix}_{\text{stat}} 
\pm 0.4_{\text{syst}}
\right)\times 10^{-4} \\\\
\text{Im}(\omega)=
\left(-1.7 
\begin{smallmatrix}+3.3\\ -3.0\end{smallmatrix}_{\text{stat}} 
\pm 1.2_{\text{syst}}
\right)\times 10^{-4}
\end{matrix} 
\right\}
\quad 
|\omega|<1.0\times 10^{-3}\ \text{at }\ 95\% \text{ C.L.} 
\end{equation}

At least one order of magnitude improvement is expected with KLOE-2 at the upgraded DA$\Phi$NE.

\section{Conclusion}
Separate tests of the Discrete Symmetries CP, T, CPT are being performed in the meson factories. These studies are made possible thanks to the spectacular quantum properties of EPR entangled states. The appropriate preparation and detection of the initial and final states in meson transitions defined by Flavour-CP eigenstates decay channels are based on Entanglement and the use of the two decays as Filtering Measurements.

\section*{Acknowledgements}
 I would like to thank Pawel Moskal and the Organising Committee for the superb atmosphere of the Conference. This research has been supported by MINECO and Generalitat Valenciana Projects FPA 2011-23596 and GVPROMETEO II 2013-017 and by Severo Ochoa Excellence Centre Project SEV 2014-0398.

\section*{References}
\begin{itemize}
\item [[1]]Cabibbo N, Phys.Rev.Lett. 10, 531 (1963).
\\Kobayashi M and Maskawa T, Prog.Theor.Phys. 49, 652 (1973).
\item[[2]]Antonelli M, Asner D M, Bauer D A, Becher T G, Beneke M, et al., Phys.Rept. 494, 197(2010).
\item[[3]]Aubert B et al. (BABAR Collaboration), Nucl.Instrum.Meth.A, 615(2013).
\\B. Aubert et al. (BABAR Collaboration), Nucl.Instrum.Meth.A 479, 1 (2002).
\item[[4]]Abashian A et al., Nucl.Instrum.Meth.A 479, 117 (2002).
\item[[5]]The LHCb Collaboration, JINST 3, S008 005(2008).
\item[[6]]Gershon T, Nir Y, in RPP, Chin. Phys. 38, Number 9 (2014).
\item[[7]]R.Aaij et al., LHCb Collaboration, arXiv:1407.6211.
\item[[8]]LHCb Collaboration, Aaij R et al., Phys. Rev. Lett. 110, 028101 (2013).
\item[[9]] Descotes-Genon S, Hofer L, Matias J, Virto J.,J.Phys.Conf.Ser. 631 (2015) 1, 012027
\item[[10]]Banuls M C and Bernabeu J, Phys.Lett.B 464, 117 (1999).
\item[[11]]Banuls M C and Bernabeu J, Nucl.Phys.B 590, 19 (2000).
\item[[12]]Wolfenstein L, Int.J.Mod.Phys.E 8, 501 (1999).
\item[[13]]Quinn H R, J.Phys.Conf.Ser. 171, 011001 (2009).
\item[[14]]Bernabeu J, Martinez-Vidal F, and Villanueva-Perez P, JHEP 1208, 064 (2012).
\item[[15]]Einstein A, Podolsky B, and Rosen N, Phys.Rev. 47, 777(1935).
\item[[16]]Reid M D, Drummond P D, Bowen W P, Cavalcanti E G, Lam P K, et al., Rev.Mod.Phys. 81,1727 (2009).
\item[[17]]BABAR Collaboration, Lees J L et al., Phys. Rev. Lett. 109(2012)211801.
\item[[18]]Carossi R et al., Phys. Lett. B 237(1990)303.
\item[[19]] Lueders G, Ann. Phys. (NY) 2, 1 (1957); Pauli W, in Niels Bohr and the development of physics,Pergamon, London, 1955, p. 30;Bell J S, Proc. R. Soc. London A 231, 479 (1955)
\item[[20]]Angelopoulos A, CPLEAR collaboration, Phys. Lett. B444, 52 (1998);Abouzaid E et al, Phys. Rev. D83, 092001 (2011)
\item[[21]]Banuls M C and Bernabeu J, JHEP 9906:032(1999).
\item[[22]]Bernabeu J, Mavromatos N E and Papavassiliou J, Phys. Rev.Lett. 92:131601(2004).
\item[[23]]Di Domenico A, 0904.1976 (2009).
\end{itemize}


\end{document}